\documentclass[twocolumn,english,pra]{revtex4}
\usepackage[T1]{fontenc}
\usepackage[latin1]{inputenc}
\usepackage{amsmath}
\usepackage{graphicx}
\usepackage{amssymb}

\makeatletter
\usepackage{graphicx}
\usepackage{graphicx}

\usepackage{babel}
\makeatother
\begin{document}

\title{Coherent Destruction of Photon Emission from a Single Molecule Source:
A Renormalization Group Approach}

\author{Igor Rozhkov, E. Barkai}

\affiliation{Department of Chemistry and Biochemistry, Notre Dame University,
Notre Dame, IN 46556}

\begin{abstract}
The photon emission from a single molecule driven simultaneously by
a laser and a slow electric radio frequency (rf) field is studied.
We use a non-Hermitian Hamiltonian approach which accounts for the
radiative decay of a two level system modeling a single molecule source.
We apply the renormalization group method for differential equations
to obtain long time solution of the corresponding Schrödinger equation,
which allows us to calculate the average waiting time for the first
photon emission. Then, we analyze the conditions for suppression and
enhancement of photon emission in this dissipative two-level system.
In particular we derive a transcendental equation, which yields the
non-trivial rf field control parameters, for which enhancement and
suppression of photon emission occurs. For finite values of radiative
decay rate an abrupt transition from the molecule's localization in
its ground state to delocalization is found for certain critical values
of the rf field parameters. Our results are shown to be in agreement
with the available experiments {[}Ch. Brunel \emph{et al}, Phys. Rev.
Lett. 81, 2679 (1998){]}. 
\end{abstract}
\maketitle

\section{Introduction}

Pump-probe experimental setup, used and studied in optical spectroscopy
for years (see, for example, \cite{pumpprobe}) plays a significant
role in fluorescence spectroscopy of single molecules. When a monochromatic
off-resonant electromagnetic field is added to already present near-resonant
one, a single molecule exhibits a great variety of quantum optical
effects. Just a few to be mentioned in this connection, are the light-shift
of the zero-phonon line, Autler-Townes splitting \cite{Pump-probe1},
multiphoton resonances \cite{Pump-probe2} and rf Rabi resonances
\cite{Orrit}. The current interest in this field, within the context
of single molecule spectroscopy, stems from the ability to use the
slow rf field to manipulate the single molecule, in such a way that
a single photon or a few photons are emitted per rf period. Such single
photon spectroscopy, is likely to be the most delicate measurement
one can perform on a molecule using optics. Furthermore, such an approach
might be useful for quantum cryptology, and quantum computing.

Possible frameworks for the analysis of existing low temperature optical
experiments on fluorescence spectra of a single molecule \cite{OrritReview}
are provided by a well known two-level atom picture. Among these frameworks
is a description of photon emission by a single molecule via optical
Bloch equations. This approach has a long history in quantum optics,
and in studies of nuclear magnetic resonance and electron-spin resonance
\cite{Cohen-Tannoudji}. Quite frequently, however, one has to settle
for the numerical solution to compare Bloch equations description
of a single molecule with laboratory measurements. 

Another common analytical tool in this context is the quantum jump
approach \cite{QuantumJump}. It treats the two-level system (TLS)
emitting photons as an open or dissipative system where the time evolution
of the system is governed by the Schrödinger equation with an effective
non-Hermitian Hamiltonian operator. The system evolves from its ground
state $\left|g\right\rangle $ to a superposition of excited state
$\left|e\right\rangle $ and state $\left|g\right\rangle $, and jumps
back into state $\left|g\right\rangle $ emitting a photon. Then the
process starts over again. It is possible to characterize such jumps
statistically, and therefore to get the statistics of photon emission
by single molecule source. 

The objective of this paper is to study the average waiting time of
the emission event by a SM interacting with laser and off-resonant
rf field. Using the renormalization group (RG) method for differential
equations \cite{ChenGoldenfeldOono} we carry out the perturbative
solution of time dependent Schrödinger equation with effective quantum
jump Hamiltonian. Our goal is to obtain the conditions for enhancement
and suppression of photon emission, as the rf field control parameters
are changed. As discussed in Ref. \cite{MakarovandMetiu}, this problem
is related to the problem of coherent destruction of tunneling (CDT)
for a particle in a double well potential \cite{Grossmann1}. 

From CDT studies it is known that specific rf field parameters may
cause particle's localization in one of the wells \cite{Grossmann1},
when the transition probability between the two states of the system
is made vanishingly small. Such a suppression or coherent destruction
of tunneling \cite{Grossmann1} corresponds to a suppression of emission
in a SM \cite{MakarovandMetiu}. However, the SM system is fundamentally
different then an electron in a double well potential, since the SM
emits photons. Therefore, we are left with the task of establishing
the influence of radiative life time of the molecule on the existing
CDT picture.

This is done in Section II, where we present the RG analysis for the
equations governing the evolution of the dissipative TLS. In Section
III we derive the criteria for localization of the system in its ground
state for the cases of small and finite radiative decay rate and discuss
the transition from localization to delocalization as well as critical
behavior of the spectrum. The conclusions are given in Section IV.
A brief summary of our results was given in \cite{PRL}.

Finally, in connection to the fluorescence peak suppression condition
it is worth mentioning the well known result from the theory of dressed
states in quantum optics. The intensity of the resonances of the atom
dressed by the rf photons is given by $J_{l}\left(\xi\right)^{2}$
($l=0,1,2\dots$), where the modulation index $\xi$ is equal to the
inner product of systems dipole moment and external field in units
of driving frequency \cite{Orrit,Cohen-Tannoudji} (see details below).
Therefore, the weight of the resonance can be made infinitesimal,
by tuning the modulation index to one of the zeroes of the corresponding
Bessel function. For the connection between the theory of dressed
states and perturbation theory used in the present paper see Ref.
\cite{Frasca2}. Note that the $l=0$ case is precisely the CDT condition
\cite{Grossmann1}. In this paper we find a new condition for the
peak suppression which takes into consideration the radiative part
of the Hamiltonian and modifies {}``zeros of Bessel function'' condition,
well known in quantum tunneling and quantum optics.

\section{RG analysis of the Schrödinger equation for a single molecule}

For a molecule with a ground electronic state $\left|g\right\rangle $
and an excited electronic state $\left|e\right\rangle $, in the presence
of a laser and a slower rf fields, the Schrödinger equation has the
form\[
i\frac{\partial}{\partial t}\left|\Psi\left(t\right)\right\rangle =H\left(t\right)\left|\Psi\left(t\right)\right\rangle ,\]
\begin{align*}
H\left(t\right)= & V_{g}\cos\omega_{rf}t\left|g\right\rangle \left\langle g\right|+\frac{\Omega}{2}\left(\left|g\right\rangle \left\langle e\right|+\left|e\right\rangle \left\langle g\right|\right)\\
+ & \left(V_{e}\cos\omega_{rf}t-i\frac{\Gamma}{2}+\delta\right)\left|e\right\rangle \left\langle e\right|.\end{align*}
 Here $\omega_{rf}$ is rf field frequency, $\delta$ is the laser
detuning, $V_{e,g}=\boldsymbol\mu_{e,g}\cdot\boldsymbol E_{rf}$,
$\Omega=\boldsymbol\mu_{eg}\cdot\mathbf{E}$ (the Rabi frequency),
where $\boldsymbol\mu_{e,g}$ are permanent dipole moments of the
molecule in states $\left|e\right\rangle $ and $\left|g\right\rangle $,
$\boldsymbol\mu_{eg}$ is the transition dipole, and $\boldsymbol E$
and $\boldsymbol E_{rf}$ stand for the amplitudes of the laser and
the rf fields respectively. The emission event is taken into account
by adding a non-Hermitian part $-\left(i\Gamma/2\right)\left|e\right\rangle \left\langle e\right|$
to the TLS Hamiltonian \cite{Cohen-Tannoudji},\begin{equation}
i\frac{\partial}{\partial t}\left|\Psi\left(t\right)\right\rangle =\left(H\left(t\right)-\left(i\Gamma/2\right)\left|e\right\rangle \left\langle e\right|\right)\left|\Psi\left(t\right)\right\rangle ,\label{SrodingerEqn}\end{equation}
 where $\Gamma$ is the rate of radiative decay, also referred to
as damping coefficient. We are interested in the evolution of $\left|\Psi\left(t\right)\right\rangle =\Psi_{g}\left(t\right)\left|g\right\rangle +\Psi_{e}\left(t\right)\left|e\right\rangle $,
or, to be precise in the evolution of the survival probability\[
P_{0}\left(t\right)=\left\langle \Psi\left(t\right)\right.\left|\Psi\left(t\right)\right\rangle ,\]
 i.e. the probability of no emission event to occur in the time interval
between $0$ and $t$. Note, that in Eq. (\ref{SrodingerEqn}) the
TLS Hamiltonian $H\left(t\right)$ is taken in the rotating wave approximation
for the fast on-resonant laser field, but not for the slow off-resonant
rf field. 

From the survival probability the average waiting time $\left\langle \tau\right\rangle $
for emission event \cite{MakarovandMetiu,QuantumJump} is readily
obtained, according to\begin{equation}
\left\langle \tau\right\rangle =\int_{0}^{\infty}P_{0}\left(t\right)dt=\int_{0}^{\infty}\left(\left|\Psi_{g}\left(t\right)\right|^{2}+\left|\Psi_{e}\left(t\right)\right|^{2}\right)dt.\label{tau1}\end{equation}
 This quantity yields the average waiting time for the first emission
when $\left|\Psi\left(0\right)\right\rangle =\left|g\right\rangle $,
and the phase of the rf field (Eq. (\ref{SrodingerEqn})) is initially
equal to zero. It can serve as an estimate for the mean time between
successive emissions, provided we can neglect the effect of the phase
of the rf field. When the Rabi frequency is small enough, such an
effect can be minimal, because the excited state population is proportional
to $\Omega^{2}$, when only laser field is present. 

In what follows\textbf{,} we discuss two cases: $\Gamma/\omega_{rf}\ll1$,
$\Omega/\omega_{rf}\ll1$ and $\Omega/\omega_{rf}\ll1$, $\Gamma/\omega_{rf}\sim1$.
The first one, allows the comparison with fluorescence measurements
carried out in Ref. \cite{Orrit} and illustrates the localization
condition of a single molecule in its ground state, induced by the
rf field. In the second case the localization gets destroyed with
the increase in dimensionless damping coefficient $\Gamma/\omega_{rf}$,
and we are able to follow the transition from localization to delocalization. 

We should also note, that Eq. (\ref{SrodingerEqn}) has time-dependent
coefficients, and no closed form solution is known for a general set
of laser and rf fields parameters. The simplification introduced by
choosing a small Rabi frequency in Eq. (\ref{SrodingerEqn}) ($\Omega/\omega_{rf}\ll1$)
leads to the possibility of perturbative calculation of $\left|\Psi\left(t\right)\right\rangle $.

\subsection{RG analysis in the limit of small radiative decay rate}

First of all we select the time scale of our problem and nondimensionalize
all its parameters. This can be done as follows: $t\mapsto\omega_{rf}t$,
$V_{e,g}\mapsto V_{e,g}/\omega_{rf}$, $\delta\mapsto\delta/\omega_{rf}$,
$\Gamma\mapsto\Gamma/\omega_{rf}$, $\Omega\mapsto\Omega/\omega_{rf}$.
By referring to the Rabi frequency as being small we mean, that it
is small compared to the frequency of the rf field. We can further
simplify our analysis of Eq. (\ref{SrodingerEqn}) by assuming the
system dissipation to be small as well. More explicitly, $\Gamma\mapsto\epsilon\Gamma$,
$\Omega\mapsto\epsilon\Omega$ ($\epsilon\ll1$). 

Next step is the change of variables according to: $\Psi_{g}\left(t\right)=c_{g}\left(t\right)\textrm{exp}\left\{ -iV_{g}\sin t\right\} $,
$\Psi_{e}\left(t\right)=c_{e}\left(t\right)\textrm{exp}\left\{ -iV_{e}\sin t-i\delta t\right\} $,
to prepare for the integration in Eq. (\ref{SrodingerEqn}). Then,
we make use of the identity \cite{Abramowitz}\begin{equation}
\textrm{exp}\left\{ iz\sin t\right\} =\sum_{k=-\infty}^{\infty}\textrm{e}^{ikt}J_{k}\left(z\right),\label{identity}\end{equation}
 and the original system (Eq. (\ref{SrodingerEqn})) reduces to the
following set of coupled ordinary differential equations:\[
\frac{d}{dt}c_{g}=-i\epsilon c_{e}\frac{\Omega}{2}\sum_{k=-\infty}^{\infty}\textrm{e}^{-ikt-i\delta t}J_{k}\left(\xi\right),\]
\begin{equation}
\frac{d}{dt}c_{e}{\displaystyle =-i\epsilon c_{g}\frac{\Omega}{2}\sum_{k=-\infty}^{\infty}\textrm{e}^{ikt+i\delta t}J_{k}\left(\xi\right)}-\frac{\epsilon\Gamma}{2}c_{e}\left(t\right),\label{systemEq2}\end{equation}
 where we introduced the modulation index $\xi=V_{e}-V_{g}$. 

We now turn to the perturbation analysis, with the objective to get
an asymptotic solution of Eqs. (\ref{systemEq2}) which would be valid
globally in time. To do that we apply the RG method introduced in
\cite{ChenGoldenfeldOono}. Unlike conventional singular perturbation
methods \cite{Nayfeh}, the RG procedure starts with the construction
of the naive expansion. It takes the form of a power series in the
small parameter $\epsilon$:\begin{equation}
C\left(t\right)=c^{\left(0\right)}\left(t\right)+\epsilon c^{\left(1\right)}\left(t\right)+\epsilon^{2}c^{\left(2\right)}\left(t\right)+O\left(\epsilon^{3}\right).\label{naive}\end{equation}
 The naive solution is then obtained by substituting $C\left(t\right)$
into Eq. (\ref{systemEq2}) and solving for the gauge functions $c^{\left(n\right)}\left(t\right)$
at each order $\epsilon^{n}$. It is assumed, that the gauge functions
are $O\left(1\right)$ for all $t$. However, this is almost always
not the case. Usually, for large enough times one or several $c^{\left(n\right)}\left(t\right)$
become greater than their predecessors $c^{\left(m\right)}\left(t\right)$
($m<n$). Then, the naive expansion becomes nonuniform and cannot
be used as an approximate solution. The terms causing nonuniformity
are called secular terms \cite{Nayfeh}. The RG procedure regularizes
the naive expansion by identifying the secular terms and eliminating
them. Although, there exist a bulk of literature on RG method and
its applications \cite{Paquette}, we outline its key steps and details
in Appendix A. 

In the naive expansion (Eq. (\ref{naive})), which we now substitute
into Eq. (\ref{systemEq2}), all the gauge functions are two-component
vectors $c^{\left(n\right)}=\left(c_{g}^{\left(n\right)},c_{e}^{\left(n\right)}\right)^{T}$.
At $O\left(\epsilon^{0}\right)$, the solution of Eq. (\ref{systemEq2})
is a constant vector $c^{\left(0\right)}=\left(A,B\right)^{T}$. One
can see, that at $O\left(\epsilon\right)$ this leads to the linearly
growing terms (secular terms) at integer values of detuning: $\delta=l$.
Thus, to be able to treat other values of detuning simultaneously
with the integer ones, we set $\delta=-l+\epsilon\sigma$, where $\sigma=O\left(1\right)$
is the {}``detuning of the detuning'' and proceed. The solution
we are about to obtain describes the vicinity of fluorescence peak
(cf. Ref. \cite{Orrit}). The entire spectrum can be approximated
as a sum of such peaks, located at different integer values of detuning,
provided that the peaks are not too wide. Since we consider the small
dissipation and small Rabi frequency case, this assumption is secure.

We introduce a parameter $\sigma$ into Eq. (\ref{systemEq2}) to
get\begin{widetext}

\begin{equation}
\frac{d}{dt}C=-i\epsilon\left(\begin{array}{cc}
0 & \frac{\Omega}{2}J_{-l}\left(\xi\right)+\frac{\Omega}{2}{\displaystyle \sum_{k\ne l}}\textrm{e}^{-ikt-ilt}J_{k}\left(\xi\right)\\
\frac{\Omega}{2}\textrm{e}^{it\epsilon\sigma}J_{-l}\left(\xi\right)+\frac{\Omega}{2}{\displaystyle \sum_{k\ne l}}\textrm{e}^{ikt+ilt}J_{k}\left(\xi\right) & -i\frac{\Gamma}{2}+\sigma\end{array}\right)C,\label{Csystemdetun1}\end{equation}
\end{widetext} with $C=\left(c_{g},c_{e}\right)^{T}$, and $C\left(0\right)=\left(1,0\right)^{T}$.
From Eq. (\ref{Csystemdetun1}) the result to $O\left(\epsilon\right)$
reads\[
C=\left(\begin{array}{c}
A-\frac{i\epsilon B\Omega}{2}J_{-l}\left(\xi\right)\left(t-t_{0}\right)\\
B-\frac{i\epsilon A\Omega}{2}J_{-l}\left(\xi\right)\left(t-t_{0}\right)-\left(\epsilon\frac{\Gamma}{2}+i\sigma\right)B\left(t-t_{0}\right)\end{array}\right),\]
 where $t_{0}$ is an arbitrary time scale used to regularize the
naive expansion (See Appendix A). In the $O\left(\epsilon\right)$
solution we dropped all the terms other than secular, due to the irrelevance
of such terms for the next steps of the procedure. The RG equations
are obtained by imposing $dC/dt_{0}=0$ at $t_{0}=t.$ We have:\begin{equation}
\frac{d}{dt}\left(\begin{array}{c}
A_{R}\\
B_{R}\end{array}\right)=-i\epsilon\left(\begin{array}{cc}
0 & \frac{\Omega}{2}J_{-l}\left(\xi\right)\\
\frac{\Omega}{2}J_{-l}\left(\xi\right) & -i\frac{\Gamma}{2}+\sigma\end{array}\right)\left(\begin{array}{c}
A_{R}\\
B_{R}\end{array}\right),\label{RGdetun1}\end{equation}
 where subscript {}``\emph{R}'' indicates, that we are solving for
the renormalized values of $A$ and $B$. Finally, solving Eq. (\ref{RGdetun1})
with the initial conditions $C\left(0\right)=\left(1,0\right)^{T}$
we obtain the following renormalized zero order solution $C\left(t\right)=\left(A_{R}\left(t\right),B_{R}\left(t\right)\right)^{T}+O\left(\epsilon\right)$:\begin{equation}
A_{R}\left(t\right)=\frac{\lambda_{1}\exp\left\{ \epsilon\lambda_{2}t\right\} -\lambda_{2}\exp\left\{ \epsilon\lambda_{1}t\right\} }{\lambda_{1}-\lambda_{2}},\label{Adetun1}\end{equation}
\begin{equation}
B_{R}\left(t\right)=\frac{i\Delta_{l}}{\lambda_{1}-\lambda_{2}}\left(\exp\left\{ \epsilon\lambda_{2}t\right\} -\exp\left\{ \epsilon\lambda_{1}t\right\} \right),\label{Bdetun1}\end{equation}
 \[
\lambda_{1,2}=-\frac{\Gamma}{4}-\frac{i\sigma}{2}\pm\frac{1}{2}\sqrt{\Gamma^{2}-16\Delta_{l}^{2}-4\sigma^{2}-4i\sigma\Gamma}.\]
 Here, by {}``$\pm$'', we specified two different square roots,
and \[
\Delta_{l}=\frac{J_{-l}\left(\xi\right)\Omega}{2}.\]
We can now set $\epsilon=1$, as we do not need it anymore. 

After the integration in Eq. (\ref{tau1}), it can be shown, that
$\left\langle \tau\right\rangle _{l}$, the average waiting time,
when the detuning is close to the integer value $l$, is given by\[
\left\langle \tau\right\rangle _{l}=\frac{\Gamma^{2}+8\Delta_{l}^{2}+4\sigma^{2}}{4\Gamma\Delta_{l}^{2}}.\]
 Then, for arbitrary laser detuning $\delta$, the average waiting
time, can be approximated as\begin{equation}
\left\langle \tau\right\rangle \approx{\displaystyle \sum_{l}}\frac{\Gamma^{2}+8\Delta_{l}^{2}+4\sigma^{2}}{4\Gamma\Delta_{l}^{2}}.\label{fluorescence1}\end{equation}
 The equality sign in Eq. (\ref{fluorescence1}) indicates the approximate
nature of this formula, which we mentioned above. The first thing
that catches attention is the presence of Bessel functions in the
denominators of each term in a sum. This opens the possibility to
suppress the fluorescence peak (to infinitely extend waiting time),
and to localize the single molecule in its ground state. For integer
values of the detuning $\delta=l$ (in units of $\omega_{rf}$) we
can always select parameters of the rf field, i.e. modulation index
$\xi$, so that $J_{\delta}\left(\xi\right)=0$.

\begin{figure}
\begin{center}\includegraphics{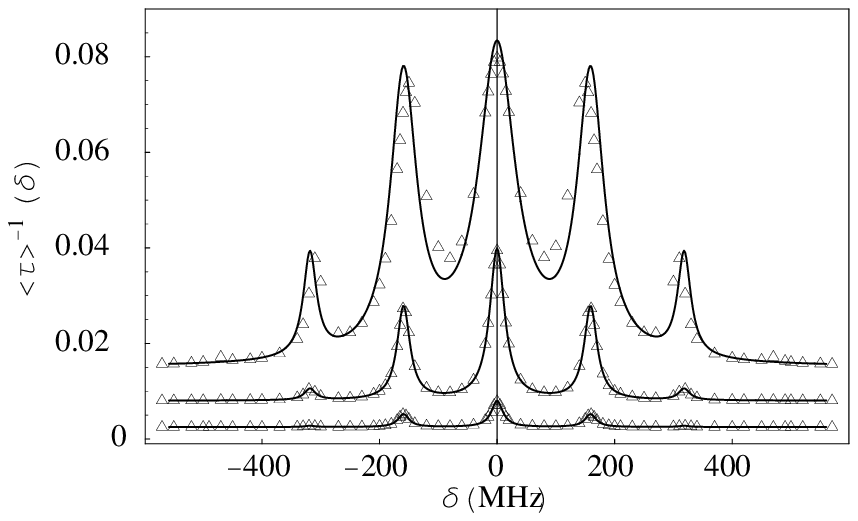}\end{center}

\begin{center}\includegraphics{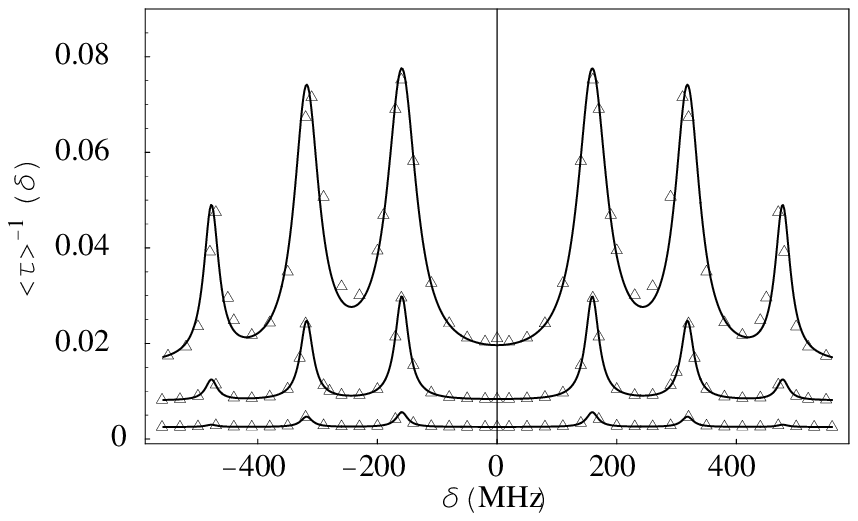}\end{center}

\caption{\label{OrritCase} Comparison of the predictions for $\left\langle \tau\right\rangle ^{-1}\left(\delta\right)$
given by Eq. (\ref{fluorescence1}) (solid lines) with the excited
state population calculated from Bloch equations (empty triangles)
(Ref. \cite{Orrit}) for various values of Rabi frequencies (from
bottom to the top): $\Omega=0.29;0.9;3.2$ in units of $\Gamma$.
Parameters are taken from experiment of Ref. \cite{Orrit}: $\omega_{rf}/2\pi=140$
MHz, (a) $\xi=1.14$; (b) $\xi$ is the first root of $J_{0}\left(\xi\right)=0$,
$\Gamma/2\pi=20$ MHz. The results are arbitrarily shifted vertically
for transparency, therefore, units on vertical axis are arbitrary.
Our predictions in (a) are in accord with experiments \cite{Orrit}.
In (b) we observe the suppression of the fluorescence for $\delta=0$.}
\end{figure}

In Fig. \ref{OrritCase}a, we compare the RG result (Eq. (\ref{fluorescence1}))
with the numerical solution of Bloch equations, previously done in
Ref. \cite{Orrit} in order to match it with their experimental results.
We repeat numerical simulations of Ref. \cite{Orrit} for the same
set of system parameters and calculate the fluorescence spectra. The
prediction of Eq. (\ref{fluorescence1}) is in a good agreement with
the numerical solution for the three smallest values of Rabi frequency
(Fig. \ref{OrritCase}). The slight discrepancy observed for the highest
value of Rabi frequency (top graph in Fig. \ref{OrritCase}a) is due
to the fact that perturbation parameter $\epsilon\approx0.46$ in
this case is rather close to $1$. In particular, the RG solution
(to this order) does not capture the shifts in the fluorescence peaks
position, found in experiments for $\Omega\geq3.2\Gamma$ \cite{Orrit}.
In Fig. \ref{OrritCase}b we demonstrate the outcome of changing the
value of modulation index $\xi$ from $1.14$ of Ref. \cite{Orrit}
to the first root of zeroth order Bessel function ($\xi\approx2.4048$).
We can see that the criterion for destruction of emission ($J_{\delta}\left(\xi\right)=0$)
is supported by numerical solution of Bloch equations, as the central
peak ($\delta=0$) gets suppressed for the latter value of $\xi$. 

\begin{figure}
\begin{center}\includegraphics{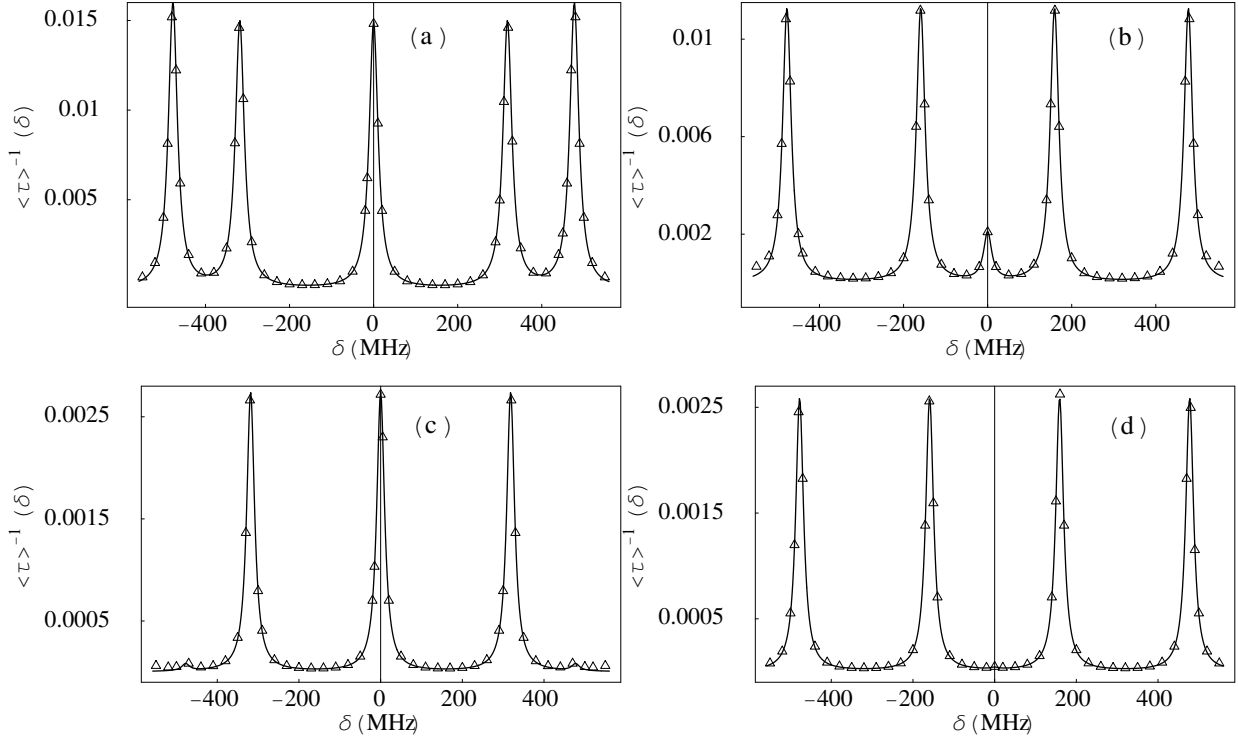}\end{center}

\caption{\label{differentksi} Suppression of the fluorescence peaks for different
values of modulation index. Here $\Omega=0.9\Gamma$, $\Gamma/2\pi=20$
MHz (parameters of Ref. \cite{Orrit}). (a) $\xi$ is the first root
of $J_{1}\left(\xi\right)=0$; (b) $\xi$ is the first root of $J_{2}\left(\xi\right)=0$;
(c) $\xi$ is the eighth root of $J_{1}\left(\xi\right)=0$; (d) $\xi$
is the eighth root of $J_{2}\left(\xi\right)=0$. }
\end{figure}

Now we further explore the emission suppression case. Keeping the
same parameters as in Ref. \cite{Orrit} we change value of $\xi$
to make the side peaks disappear. The results are presented in Fig.
\ref{differentksi}. We also observe, that for sufficiently large
$\xi$, i.e. large number of the Bessel function root, we can suppress
more than two side peaks, corresponding to positive and negative integer
detunings. Taking $\xi$ to be the $8$th root of $J_{1}\left(\xi\right)$
we eliminate the third peaks, while taking it to be the $8$th root
of $J_{2}\left(\xi\right)$ we eliminate the central peak ($\delta=0$)
as well. This can be explained by the behavior of zeroes of Bessel
functions, which fall densely on the real axis, as the argument increases.
For the large modulation index one expects the entire spectrum to
be suppressed \cite{Triggered}.

\subsection{Average waiting time for the case of finite decay rate}

The evolution of survival probability alters significantly when the
dissipation coefficient $\Gamma$ is larger, or of the same order
as the driving frequency $\omega_{rf}$. This becomes clear from the
structure of the secular terms as we take on the long time asymptotic
solution for this case. We still consider Rabi frequency to be small
($\Omega\mapsto\epsilon\Omega$, $\epsilon\ll1$), but keep $\Gamma=O\left(1\right)$.
First of all, we focus on the case of zero detuning ($\delta=0$).
The Schrödinger equation then reads\begin{equation}
i\frac{\partial}{\partial t}\Psi=\left(\begin{array}{cc}
V_{g}\cos t & \epsilon\Omega/2\\
\epsilon\Omega/2 & V_{e}\cos t-i\Gamma/2\end{array}\right)\Psi.\label{system}\end{equation}
 The transformation to {}``amplitude'' variables is almost the same
as before: $\Psi_{g}\left(t\right)=c_{g}\left(t\right)\textrm{e}^{-iV_{g}\sin t},$
$\Psi_{e}\left(t\right)=c_{e}\left(t\right)\textrm{e}^{-iV_{e}\sin t-\Gamma t/2}.$
Once again, we face a set of coupled linear ODEs:\begin{widetext}\begin{equation}
\frac{d}{dt}C=-i\frac{\Omega}{2}\epsilon\left(\begin{array}{cc}
0 & \sum_{k=-\infty}^{\infty}\textrm{e}^{-ikt-\frac{\Gamma}{2}t}J_{k}\left(\xi\right)\\
\sum_{k=-\infty}^{\infty}\textrm{e}^{ikt+\frac{\Gamma}{2}t}J_{k}\left(\xi\right) & 0\end{array}\right)C.\label{Csystem}\end{equation}
\end{widetext}Proceeding with the naive solution, we observe that the secular terms
do not appear in this expansion at $O\left(\epsilon\right)$, as they
did in the case of weak dissipation. 

\begin{figure}
\begin{center}\includegraphics{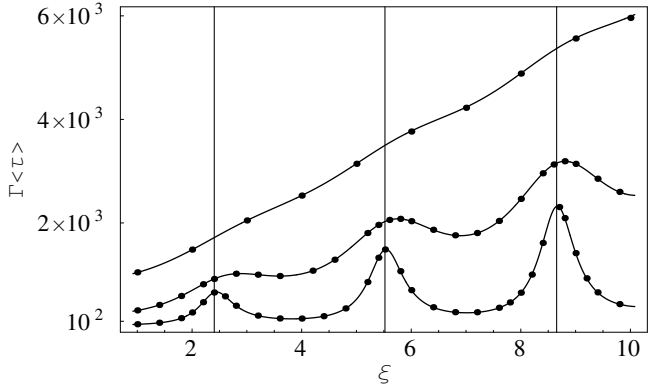}\end{center}

\caption{\label{MTau} Scaled mean time between emission $\Gamma\left\langle \tau\right\rangle $
is plotted as a function of modulation index $\xi$ for the case of
zero detuning. The solid line represents RG prediction of Eq. (\ref{tbar3}),
dots correspond to the numerical solution of Eq. (\ref{SrodingerEqn}).
Grid-lines indicate the positions of zeroes of $J_{0}\left(\xi\right)$.
The Rabi frequency is $\Omega=0.1$. The radiative decay rates are
$\Gamma=0.5$; $1.5$; $3.0$ (from top to bottom). We notice the
shifts in maximum of $\left\langle \tau\right\rangle $ relative the
zeroes of $J_{0}\left(\xi\right)$.}
\end{figure}

At $O\left(1\right)$, the solution is a constant vector $c^{\left(0\right)}=\left(A,B\right)^{T}$.
Substitution of this result into the $O\left(\epsilon\right)$ equations
yields:\[
c_{g}^{\left(1\right)}=i\Omega B\sum_{k=-\infty}^{\infty}J_{k}\left(\xi\right)\left(\textrm{e}^{-\Gamma t/2-ikt}-1\right)\frac{\Gamma-2ik}{\Gamma^{2}+4k^{2}},\]
 \[
c_{e}^{\left(1\right)}=i\Omega A\sum_{k=-\infty}^{\infty}J_{k}\left(\xi\right)\left(\textrm{e}^{\Gamma t/2+ikt}-1\right)\frac{\Gamma-2ik}{\Gamma^{2}+4k^{2}},\]
 which, in turn, leads to the following $O\left(\epsilon^{2}\right)$
equations:\begin{align*}
\frac{d}{dt}c_{g}^{\left(2\right)} & =\frac{\Omega^{2}}{2}A\sum_{k,l=-\infty}^{\infty}J_{k}\left(\xi\right)J_{l}\left(\xi\right)\textrm{e}^{-\Gamma t/2-ikt}\\
 & \times\left(\textrm{e}^{\Gamma t/2+ilt}-1\right)\frac{\Gamma-2il}{\Gamma^{2}+4l^{2}},\end{align*}
\begin{align*}
\frac{d}{dt}c_{e}^{\left(2\right)} & =\frac{\Omega^{2}}{2}B\sum_{k,l=-\infty}^{\infty}J_{k}\left(\xi\right)J_{l}\left(\xi\right)\textrm{e}^{\Gamma t/2+ikt}\\
 & \times\left(1-\textrm{e}^{-\Gamma t/2-ilt}\right)\frac{\Gamma-2il}{\Gamma^{2}+4l^{2}},\end{align*}
 Further integration produces secular terms (for $k=l$)\[
c_{g}^{\left(2\right)}=-A\frac{t\Omega^{2}\Gamma}{2}\sum_{k=-\infty}^{\infty}\frac{J_{k}^{2}\left(\xi\right)}{\Gamma^{2}+4k^{2}}+NST,\]
 \[
c_{e}^{\left(2\right)}=-B\frac{t\Omega^{2}\Gamma}{2}\sum_{k=-\infty}^{\infty}\frac{J_{k}^{2}\left(\xi\right)}{\Gamma^{2}+4k^{2}}+NST,\]
 where {}``NST'' stands for nonsecular contribution to $O\left(\epsilon^{2}\right)$
solutions. The RG equations (to order $O\left(\epsilon^{3}\right)$)
for the regularized {}``constants'' $A$ and $B$ are:\[
\frac{dA_{R}}{dt}=-A_{R}\frac{\epsilon^{2}\Omega^{2}\Gamma}{2}\sum_{k=-\infty}^{\infty}\frac{J_{k}^{2}\left(\xi\right)}{\Gamma^{2}+4k^{2}},\]
 \[
\frac{dB_{R}}{dt}=-B_{R}\frac{\epsilon^{2}\Omega^{2}\Gamma}{2}\sum_{k=-\infty}^{\infty}\frac{J_{k}^{2}\left(\xi\right)}{\Gamma^{2}+4k^{2}}.\]
 Hence, to $O\left(\epsilon^{3}\right)$ $A_{R}$ and $B_{R}$ satisfy\[
A_{R}\left(t\right)=A\left(0\right)\exp\left\{ -\frac{\epsilon^{2}\Omega^{2}\Gamma t}{2}\sum_{k=-\infty}^{\infty}\frac{J_{k}^{2}\left(\xi\right)}{\Gamma^{2}+4k^{2}}\right\} ,\]
\begin{equation}
B_{R}\left(t\right)=B\left(0\right)\exp\left\{ -\frac{\epsilon^{2}\Omega^{2}\Gamma t}{2}\sum_{k=-\infty}^{\infty}\frac{J_{k}^{2}\left(\xi\right)}{\Gamma^{2}+4k^{2}}\right\} .\label{AandB}\end{equation}
 At time $t=0$ we assume that the molecule is in the ground state:
$c_{e}\left(0\right)=0$, $c_{g}\left(0\right)=1$. Applying these
initial conditions and replacing $A$ and $B$ with their renormalized
values (Eqs. (\ref{AandB})) in $c^{\left(0\right)}$ we obtain $O\left(\epsilon^{2}\right)$
perturbation results for amplitudes $c_{e}$ and $c_{g}$. Then, we
switch back to the original variables $\Psi_{e,g}\left(t\right)$
and arrive at \begin{widetext} \begin{align}
\Psi_{g} & =\textrm{e}^{-iV_{g}\sin t}\exp\left\{ -\frac{\epsilon^{2}t\Omega^{2}\Gamma}{2}\sum_{k=-\infty}^{\infty}\frac{J_{k}^{2}\left(\xi\right)}{\Gamma^{2}+4k^{2}}\right\} \nonumber \\
 & \times\left[1-\epsilon^{2}\Omega^{2}\sum_{k\ne l=-\infty}^{\infty}\frac{J_{k}\left(\xi\right)J_{l}\left(2\xi\right)}{\Gamma+2il}\left\{ \frac{i\left(1-\textrm{e}^{i\left(l-k\right)t}\right)}{l-k}-\frac{2\left(1-\textrm{e}^{-\Gamma t/2-ikt}\right)}{\Gamma+2ik}\right\} \right],\label{Psig}\end{align}
\begin{align}
\Psi_{e} & =-i\epsilon\Omega\textrm{e}^{-iV_{e}\sin t-\frac{\Gamma}{2}t}\exp\left\{ -\frac{\epsilon^{2}t\Omega^{2}\Gamma}{2}\sum_{k=-\infty}^{\infty}\frac{J_{k}^{2}\left(\xi\right)}{\Gamma^{2}+4k^{2}}\right\} \label{Psie}\\
 & \times\sum_{k=-\infty}^{\infty}J_{k}\left(\xi\right)\left(\textrm{e}^{\Gamma t/2+ikt}-1\right)\frac{\Gamma-2ik}{\Gamma^{2}+4k^{2}}.\nonumber \end{align}
\end{widetext}Then, using Eq. (\ref{tau1}) we calculate the mean waiting time $\langle\tau\rangle$;
the $O\left(\epsilon^{2}\right)$ result reads\begin{equation}
\left\langle \tau\right\rangle ^{-1}=\Gamma\Omega^{2}\sum_{k=-\infty}^{\infty}\frac{J_{k}^{2}\left(\xi\right)}{\Gamma^{2}+4k^{2}}+O\left(\frac{\Omega^{2}}{\Gamma^{2}}\right).\label{tbar3}\end{equation}
 The corresponding result for the case of nonzero detuning can be
calculated in a similar fashion. Formally, this case is different
from the one we just considered by addition of imaginary part to the
decay rate: $\Gamma\mapsto\Gamma+2i\delta$. Therefore, the general
formula for the case of arbitrary laser detuning can be written as\begin{equation}
\left\langle \tau\right\rangle ^{-1}=\Gamma\Omega^{2}\sum_{k=-\infty}^{\infty}\frac{J_{k}^{2}\left(2\xi\right)}{\Gamma^{2}+4\left(k-\delta\right)^{2}}+O\left(\frac{\Omega^{2}}{\Gamma^{2}}\right).\label{tbarnonzerodelta}\end{equation}
 The possibility of locking the system in state $\left|g\right\rangle $
for a long time, is still present. As one can see, the presence of
finite damping has only modified the corresponding condition, and
made $\left\langle \tau\right\rangle $ finite at all values of $\xi$
in contrast to the results of Eq. (\ref{fluorescence1}). In other
words, the average waiting time can be maximized for certain values
of modulation index, but the height of the peaks at the maxima positions
is now finite even at the leading order in small Rabi frequency. In
Fig. \ref{MTau} present the comparison of this prediction with the
results of numerical solution of Eq. (\ref{SrodingerEqn}). Predictions
for all three values of $\Gamma$ are adequate. 

From Fig. \ref{MTau} we observe, that when the dimensionless $\Gamma$
is not too large, the maxima in the average waiting time occur at
zeroes of $J_{\delta}\left(\xi\right)$, in accord with the results
obtained in small dissipation limit. If the decay rate is increased,
one expects the peaks to get broaden and shifted. This result does
follow from the comparison of $\Gamma=0.5$ and $\Gamma=1.0$ cases
(see Fig. \ref{MTau}). Meanwhile, the third curve in Fig. \ref{MTau},
corresponding to the case of $\Gamma=3.0$, points to a less trivial
feature in the peaks behavior. Considering the presence of well defined
maxima in $\left\langle \tau\right\rangle $ as a sign of localization
(in ground state) effect, we can refer to their disappearance as a
delocalization effect. 

The detailed study of such a transition is the purpose of the next
Section.

\section{Transition from localized to delocalized regime}

As long as the decay rate does not exceed certain threshold value,
the average waiting time $\left\langle \tau\right\rangle $, plotted
as a function of $\xi$, appears as a sequence of alternating minima
and maxima (see Fig. \ref{MTau}). Those extremums correspond to the
enhancement and reduction in photon emission rate, respectively. Setting
$d\left\langle \tau\right\rangle /d\xi=0$ and differentiating Eq.
(\ref{tbar3}) we obtain the following conditions for the extremums:\begin{align}
J_{0}\left(\xi\right) & =\frac{\Gamma^{2}}{J_{1}\left(\xi\right)}\sum_{k=1}^{\infty}\frac{J_{k}\left(\xi\right)\left(J_{k-1}\left(\xi\right)-J_{k+1}\left(\xi\right)\right)}{\Gamma^{2}+4k^{2}}\label{nun}\\
 & =G\left(\Gamma,\xi\right).\nonumber \end{align}
 Thus, for zero laser detuning, if $\Gamma$ is infinitesimal, the
$n$th maxima would be located at $\xi_{n}$, the $n$th zero of the
Bessel function $J_{0}\left(\xi\right)$ (see also Appendix A for
the derivation of corresponding condition in absence of dissipation).
When dissipation becomes noticeable, the peaks of $\left\langle \tau\right\rangle $
shift with respect to the $\xi_{n}$. Assuming that these shifts are
still small, one can estimate them perturbatively from Eq. (\ref{tbar3}).

\subsection{Condition for the emission suppression in the small shifts limit}

At the position of a maximum, we have $d\left\langle \tau\right\rangle /d\xi=0$.
Upon differentiation in Eq. (\ref{tbar3}), we use the sum rule \cite{Abramowitz}
$\sum_{k=-\infty}^{\infty}{\displaystyle J_{k}^{2}\left(\xi\right)=2{\textstyle \sum_{k=-\infty}^{\infty}}J_{k}^{2}\left(\xi\right)=1-J_{0}^{2}\left(\xi\right)}$,
which implies\[
2\sum_{k=1}^{\infty}J_{k}\left(\xi\right)\frac{dJ_{k}\left(\xi\right)}{d\xi}=-J_{0}\left(\xi\right)\frac{dJ_{0}\left(\xi\right)}{d\xi},\]
 and find, that\[
\frac{d\left\langle \tau\right\rangle }{d\xi}\propto\sum_{k=1}^{\infty}\frac{J_{k}\left(\xi\right)\frac{dJ_{k}\left(\xi\right)}{d\xi}}{\Gamma^{2}+4k^{2}}k^{2}.\]
 Then we use an identity $dJ_{k}\left(\xi\right)/d\xi=\left(J_{k-1}\left(\xi\right)-J_{k+1}\left(\xi\right)\right)/2$
and addition theorem \cite{Abramowitz} $J_{k}\left(u\pm v\right)=\sum_{l=-\infty}^{\infty}J_{k\mp l}\left(u\right)J_{l}\left(v\right)$
( $\left|v\right|<\left|u\right|$) to get\begin{widetext}\[
\frac{d\left\langle \tau\right\rangle }{d\xi}\propto\sum_{k=1}^{\infty}\frac{\sum_{l=-\infty}^{\infty}J_{k-l}\left(\xi_{n}\right)J_{l}\left(\nu_{n}\right)\sum_{l=-\infty}^{\infty}\left\{ J_{k-l-1}\left(\xi_{n}\right)-J_{k-l+1}\left(\xi_{n}\right)\right\} J_{l}\left(\nu_{n}\right)}{\Gamma^{2}+4k^{2}}k^{2},\]
\end{widetext}where $\nu_{n}$ is the shift of the $n$th peak. Next we make use
of the small $\nu_{n}$ asymptotics of Bessel functions: $J_{0}\left(\nu_{n}\right)\sim1-\nu_{n}^{2}/4$,
$J_{l}\left(\nu_{n}\right)\sim\left(\nu_{n}/2\right)^{l}$ ($l>1$).
Keeping the zeroth and the first order terms, and solving for $\nu_{n}$
we arrive at\begin{equation}
\nu_{n}=\frac{\sum_{k=1}^{\infty}\frac{k^{2}}{\Gamma^{2}+4k^{2}}J_{k}\left(\xi_{n}\right)\left\{ J_{k-1}\left(\xi_{n}\right)-J_{k+1}\left(\xi_{n}\right)\right\} }{\sum_{k=1}^{\infty}\frac{k^{2}}{\Gamma^{2}+4k^{2}}\left(P_{k}+Q_{k}\right)},\label{nun1}\end{equation}
 with\[
P_{k}=\left\{ J_{k-1}\left(\xi_{n}\right)-J_{k+1}\left(\xi_{n}\right)\right\} ^{2},\]
\[
Q_{k}=J_{k}\left(\xi_{n}\right)\left\{ J_{k-2}\left(\xi_{n}\right)-2J_{k}\left(\xi_{n}\right)+J_{k+2}\left(\xi_{n}\right)\right\} \]

\begin{figure}
\begin{center}\includegraphics{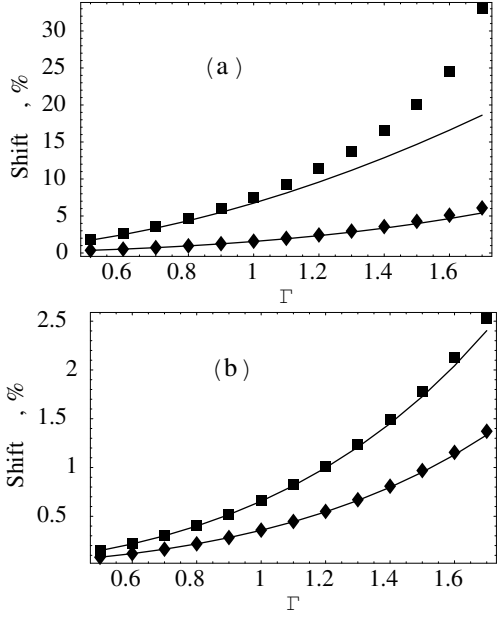}\end{center}

\caption{\label{Shifts3} Shifts in the position of maxima of $\left\langle \tau\right\rangle $
(see Fig. \ref{MTau}) with respect to (a) $\xi_{1}$ (top) and $\xi_{2}$
(bottom); (b) $\xi_{3}$ (top) and $\xi_{4}$ (bottom), plotted as
a function of $\Gamma$}
\end{figure}

For $n=1$ through $4$, we plot the relative shifts $\nu_{n}/\xi_{n}$,
given by Eq. (\ref{nun1}), as a function of damping coefficient $\Gamma$
in Fig. \ref{Shifts3} and compare them with numerical solution of
the transcendental Eq. (\ref{nun}).

It is clear that estimates given by Eq. (\ref{nun1}) predict the
shifts, which are less than $10$\%, but fail to reproduce the larger
shifts. Further analysis of extremums behavior has to be done by solving
Eq. (\ref{nun}) numerically.

\subsection{Condition for the emission suppression for arbitrary shifts}

\begin{figure}
\begin{center}\includegraphics{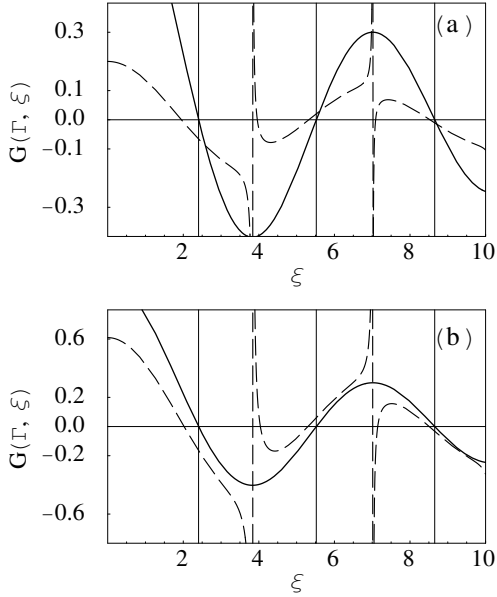}\end{center}

\caption{\label{Crossings} Illustration to the transcendental Eq. (\ref{nun}).
Bessel function $J_{0}$ (solid line) plotted as a function of $\xi$
together with the right hand side of Eq. (\ref{nun}), $G\left(\Gamma,\xi\right)$.
Radiative decay rates are (a) $\Gamma=1.0$, (b) $\Gamma=2.5$. Grid-lines
indicate zeroes of $J_{0}\left(\xi\right)$. In localization case
(a), the solutions of Eq. (\ref{nun}) are slightly shifted with respect
to $\xi_{n}$. In delocalization case (b), these solutions do not
exist since the curves do not cross.}
\end{figure}

In Fig. \ref{Crossings} we plot $G\left(\Gamma,\xi\right)$ and $J_{0}\left(\xi\right)$
versus $\xi$ to illustrate the solution of the transcendental Eq.
(\ref{nun}) for two values of $\Gamma$. When $\Gamma=1.0$ the corresponding
curves cross close to $\xi_{n}$ (Fig. \ref{Crossings}a). However,
as $\Gamma$ is increased to $2.5$, Eq. (\ref{nun}) does not have
any solutions in the vicinity of $\xi_{1}$ and $\xi_{2}$ and the
first two maxima of $\left\langle \tau\right\rangle $ disappear (Fig.
\ref{Crossings}b). We also notice, that for this value of the decay
rate, the first two minima of $\left\langle \tau\right\rangle $,
located next to the zeroes of $J_{1}\left(\xi\right)$ (Eq. (\ref{nun}))
for smaller $\Gamma$, disappear as well. 

\begin{figure}
\begin{center}\includegraphics{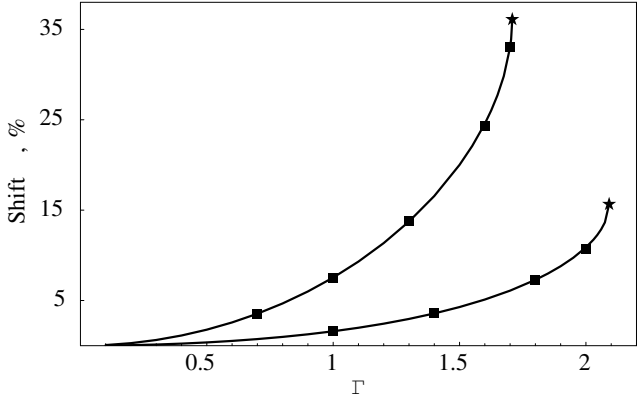}\end{center}

\caption{\label{Shifts}The shifts $\left(\xi-\xi_{n}\right)/\xi_{n}$ found
from Eq. (\ref{nun}) (solid lines) are compared to those found from
numerical solution of Eq. (\ref{SrodingerEqn}) (boxes). They are
plotted as a function of $\Gamma$ ($\Gamma/\omega_{rf}$ in original
units) for the peaks close to the first (upper curve) and the second
(lower curve) zeroes of $J_{0}$ (the first two peaks in Fig. \ref{MTau}).
Critical points are marked by stars. They indicate values of $\Gamma/\omega_{rf}$above
which the transition to delocalization occurs. }
\end{figure}

Apparently, maxima and minima of the waiting time $\langle\tau\rangle$
do not vanish gradually due to the peaks broadening. Instead, there
is a critical value of $\Gamma$ ($\Gamma/\omega_{rf}$ in original
units) for each extremum. Therefore, with an increase in dimensionless
decay rate $\Gamma/\omega_{rf}$ a single molecule undergoes the transition
from localized-like behavior to delocalized behavior, for which we
can no longer speak about maximizing average waiting time, or about
the localization of a molecule in its ground state. The transition
manifests itself as a motion of the neighboring pairs of maxima and
minima towards each other with subsequent annihilation at the same
values of critical damping $\Gamma_{cr}$ (see Fig. \ref{Crossings}).
The shifts, given by numerical solution of Eq. (\ref{nun}) are plotted
versus radiative decay rate $\Gamma$ in Fig. \ref{Shifts}. They
are in accord with the shifts obtained from the numerical solution
of the Schrödinger equation (\ref{SrodingerEqn}) for the first two
peaks of average waiting time. Slopes of both curves in Fig. \ref{Shifts}
start to diverge, as the dissipation grows, in anticipation of critical
points. For example, \textbf{$\Gamma_{cr}\simeq1.7$} for the first
peak.

\begin{figure}
\begin{center}\includegraphics{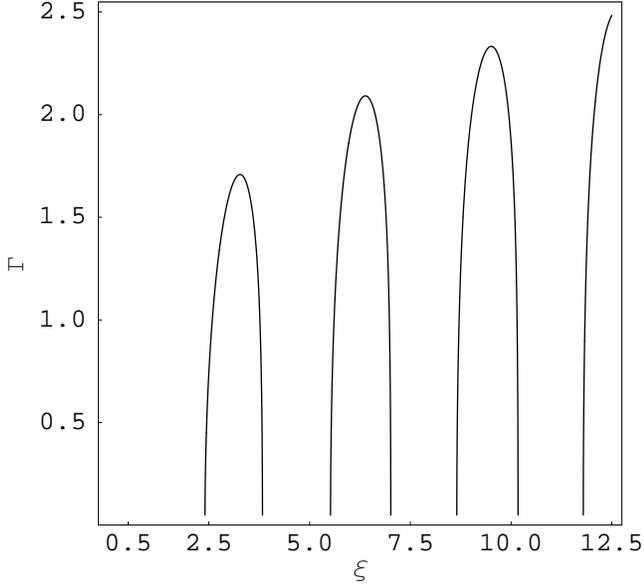}\end{center}

\caption{\label{GammaKsiplane} Graphical solution to Eq. (\ref{nun}) represented
as a solid curve in $\left(\xi,\Gamma\right)$-plane. The solution
represents the value of control parameters which maximize and minimize
$\langle\tau\rangle$. The left edge of each {}``tooth'' corresponds
to the maxima (intersecting the $\xi$-axis at the zeroes of $J_{0}\left(\xi\right)$:
$2.40483$, $5.52008$, $8.65373$, \emph{etc}.); the right edge corresponds
to the minima. The critical points (we can see three of them) are
identified as {}``teeth'' tips (cf. Fig. \ref{Crossings}).}
\end{figure}

At a critical point, the slopes of $J_{0}\left(\xi\right)$ and $G\left(\Gamma,\xi\right)$
in Fig. \ref{Crossings} coincide. For the dissipation approaching
its critical value we can use this fact and determine the asymptotic
behavior of extremums. Expansion in Taylor series of the Eq. (\ref{nun})
in the vicinity of critical point $\left(\xi_{cr},\Gamma_{cr}\right)$,
produces\[
\frac{d^{2}J_{0}\left(\xi_{cr}\right)}{d\xi^{2}}\left(\xi-\xi_{cr}\right)^{2}\sim\frac{dG\left(\Gamma_{cr},\xi_{cr}\right)}{d\Gamma}\left(\Gamma-\Gamma_{cr}\right).\]
 Here we retained only the leading terms in both sides of the equation.
Thus, the position of extremum scales with the decay rate as $\xi_{cr}-\xi\propto\left(\Gamma_{cr}-\Gamma\right)^{\beta}$,
with $\beta=1/2$. Further illustration to the transition from localization
to delocalization is provided in Fig. \ref{GammaKsiplane}, where
we display the graphical solution of Eq. (\ref{nun}) in $\left(\xi,\Gamma\right)$-plane.

\section{Discussion and conclusions}

We have shown, that a dissipative two-level system, interacting with
continuous laser field can be localized in its ground state by appropriate
choice of parameters of additional driving field. To demonstrate this
we considered a single molecule governed by a non-Hermitian Hamiltonian
in rotating wave approximation for the laser field. The added slow
rf field interacting with a molecule via dipole moments of its ground
and excited states makes the TLS Hamiltonian time dependent. Our main
quantity of interest, the average waiting time for the first emission
requires knowledge of the survival probability, which, in turn, demands
the solution of Schrödinger equation. 

Generally, this goal can be achieved only by perturbation analysis
in the case when some of the governing parameters are small. For this
reason, throughout the paper we assumed that the Rabi frequency $\Omega$
of a single molecule is small compared to the frequency of the driving
rf field $\omega_{rf}$. On the other hand, to provide a correct estimate
for the average waiting time, the resulting survival probability has
to be given by a uniformly convergent series. In this paper we overcame
this difficulty by applying the RG procedure formulated in Ref. \cite{ChenGoldenfeldOono}.
For the radiative decay rate of the excited state $\Gamma$, we considered
two limiting situations: $\Gamma/\omega_{rf}\ll1$ and $\Gamma/\omega_{rf}\sim1$.
The RG calculation for the first case generates two important results. 

(i) The fluorescence spectra of single molecule, calculated as a reciprocal
average waiting time reproduces experimental results of Ref. \cite{Orrit}. 

(ii) There is a condition for suppression of fluorescence peaks, which
resembles the criteria of coherent destruction of tunneling \cite{Grossmann1}
of electron in a double well. 

Similar procedure in the second case also produces the correct estimates
for the average waiting time $\left\langle \tau\right\rangle $, which
was verified by numerical solution of Schrödinger equation for a single
molecule. However, the localization of a molecule in its ground state
is achieved for the values of modulation index different from those
in the small $\Gamma$ limit (the well known condition $J_{0}\left(\xi\right)=0$
does not apply). Moreover, at a certain critical value of dissipation
coefficient, the neighboring minima and maxima of $\left\langle \tau\right\rangle $
annihilate, and the localization no longer persists. 

In connection with the transition from the localized behavior to the
delocalized one, we were able to estimate the shifts in the positions
of extremums of average waiting time for a given dissipation, and
to determine the critical values of $\Gamma/\omega_{rf}$ for a given
extremum. 

Finally, we remark, that similar analysis can be carried out for the
system of optical Bloch equations, which represents a standard tool
in single molecule spectroscopy, nuclear magnetic resonance and electronic
spin resonance.

\begin{acknowledgments}
This work was supported by National Science Foundation award CHE-0344930.
IR wants to thank Alexei Akolzin for valuable comments and suggestions.
\end{acknowledgments}

\appendix

\section{Renormalization group method for differential equations}

Here we illustrate basic steps of the RG method \cite{ChenGoldenfeldOono}
on two examples. The first one is a trivial linear differential equation 

\[
\partial_{t}\psi+i\omega\psi=\epsilon\psi,\]
 with $\epsilon\ll1$. Although the exact solution is known, we wish
to recover it perturbatively. Below, and everywhere in the text, the
regularization of the naive solution is achieved by analyzing the
integration constants entering the solution. These {}``constants''
play the role of bare quantities of quantum field theory and statistical
mechanics. They are treated as functions of time and are modified
(renormalized) in order to remove secular terms. 

We start by looking for a solution given as a power series in small
parameter $\epsilon$:\begin{equation}
\psi=\psi_{0}+\epsilon\psi_{1}+\epsilon^{2}\psi_{2}+O\left(\epsilon^{3}\right),\label{expansionforpsi}\end{equation}
 where $\psi_{k}$are $O\left(1\right)$ quantities. At lowest order,
we have\[
\partial_{t}\psi_{0}+i\omega\psi_{0}=0\Rightarrow\psi_{0}=A\exp\left\{ -i\omega t\right\} .\]
 We substitute this result into the equation for the first correction
to get\[
\partial_{t}\psi_{1}+i\omega\psi_{1}=A\exp\left\{ -i\omega t\right\} ,\]
 which leads to\[
\psi_{1}=At\exp\left\{ -i\omega t\right\} ,\]
as we only need a special solution of the inhomogeneous equation (the
solution to the homogeneous one is absorbed into the zeroth order
perturbation result $\psi_{0}$). The first correction grows linearly,
and after $1/\epsilon$ amount of time is elapsed $\psi_{1}$ becomes
comparable to supposedly dominating term $\psi_{0}$. Thus,\begin{equation}
\psi=A\exp\left\{ -i\omega t\right\} +\epsilon tA\exp\left\{ -i\omega t\right\} +O\left(\epsilon^{2}\right).\label{naiveexpansion}\end{equation}
 does not represent uniformly valid result. Nevertheless, according
to the RG formalism, this expansion gives us all we need for the construction
of $O\left(\epsilon^{2}\right)$ long time asymptotic expansion. 

We now assume that $A$ is not a constant, but a function of some
arbitrary time scale $t_{0}$. The renormalization consists of two
stages: first, we remove divergency to $O\left(\epsilon^{2}\right)$
by the following manipulation:\begin{align}
\psi & =A\left(1+\epsilon t_{0}\right)\exp\left\{ -i\omega t\right\} +\epsilon\left(t-t_{0}\right)A\exp\left\{ -i\omega t\right\} \nonumber \\
 & =A_{R}\left(t_{0}\right)\exp\left\{ -i\omega t\right\} +\epsilon\left(t-t_{0}\right)A_{R}\left(t_{0}\right)\exp\left\{ -i\omega t\right\} ,\label{renormalizedpsi}\end{align}
 literally {}``renormalizing'' $A$, and introducing new variable
$A_{R}$ instead. The latter expansion, is uniform in time, as $\tau$
can be chosen arbitrarily large. However, $\psi$ should not be a
function of some artificial time scale. Therefore, at the second stage,
we require\begin{equation}
\frac{\partial}{\partial t_{0}}\psi\left(t_{0}=t\right)=0\label{dpsidtau}\end{equation}
 Next, we eliminate $t_{0}$ by setting it equal to $t$ (since $t_{0}$
is arbitrary, it is at our disposal). As a result of Eq. (\ref{dpsidtau}),
we obtained differential equation (called the RG equation) governing
our former constant $A$. It reads:\[
\frac{d}{dt}A_{R}\left(t\right)=\epsilon A_{R}\left(t\right)+O\left(\epsilon^{2}\right).\]
 Finally, solving the RG equation, substituting the result into the
expansion (\ref{renormalizedpsi}), and setting $\tau=t$ we arrive
at the long time asymptotic solution uniformly valid to the next order
in $\epsilon$:\[
\psi=A\left(0\right)\exp\left\{ \epsilon t-i\omega t+O\left(\epsilon^{2}t\right)\right\} +O\left(\epsilon^{2}\right).\]
 For this simple problem it can be further shown, by repetition of
the above procedure at all orders, that $\psi=A\left(0\right)\exp\left\{ \epsilon t-i\omega t\right\} $
is actually a global solution. 

A more complicated example of Schrödinger equation for the electron
in a double well potential \cite{Grossmann1} can be also treated
using RG method. Here, our presentation is slightly different from
solution of the this problem given in Ref. \cite{Frasca2}, but it
leads to the same results and illustrates all the main steps of the
derivations in Section II. 

The TLS Hamiltonian in this case, is:\[
H=V_{g}\cos\omega t\left|g\right\rangle \left\langle g\right|+V_{e}\cos\omega t\left|e\right\rangle \left\langle e\right|+\frac{\Omega}{2}\left(\left|g\right\rangle \left\langle e\right|+\left|e\right\rangle \left\langle g\right|\right).\]
 With this Hamiltonian, the Schrödinger equation for the TLS, which
is initially in its $\left|g\right\rangle $ state results in the
following system of coupled ordinary differential equations for amplitudes
$c_{e}$ and $c_{g}$\begin{equation}
\frac{dC}{dt}=-\frac{i\Omega\epsilon}{2}\left(\begin{array}{cc}
0 & {\displaystyle \sum_{k=-\infty}^{\infty}}\textrm{e}^{-ik\omega t}J_{k}\left(\xi\right)\\
{\displaystyle \sum_{k=-\infty}^{\infty}}\textrm{e}^{ik\omega t}J_{k}\left(\xi\right) & 0\end{array}\right)C\label{Csystem}\end{equation}
 where $C=\left(c_{g},c_{e}\right)^{T}$, and $C\left(0\right)=\left(1,0\right)^{T}$
and we used Eq. (\ref{identity}). 

At $O\left(\epsilon^{0}\right)$, the solution is simply $C^{\left(0\right)}=\left(A,B\right)^{T}$,
with both $A$ and $B$ being constants. Naively, we would apply the
initial conditions at this stage. Instead, we substitute this result
into the equations for next order correction, and obtain\begin{align*}
c_{g}^{\left(1\right)} & =-i\frac{\Omega}{2}B\left\{ J_{0}\left(\xi\right)t\right.\\
 & \left.+\sum_{k=-\infty,k\ne0}^{\infty}\frac{J_{k}\left(\xi\right)\left(\textrm{e}^{-ikt}-1\right)}{k}\right\} ,\end{align*}
\begin{align*}
c_{e}^{\left(1\right)} & =-i\frac{\Omega}{2}A\left\{ J_{0}\left(\xi\right)t\right.\\
 & \left.+\sum_{k=-\infty,k\ne0}^{\infty}\frac{J_{k}\left(\xi\right)\left(1-\textrm{e}^{ikt}\right)}{k}\right\} ,\end{align*}
 The renormalized asymptotic expansions are\[
c_{g}=A_{R}-i\epsilon B_{R}\frac{\Omega}{2}\left(t-\tau\right)J_{0}\left(\xi\right)+NST+O\left(\epsilon^{2}\right),\]
 \[
c_{e}=B_{R}-i\epsilon A_{R}\frac{\Omega}{2}\left(t-\tau\right)J_{0}\left(\xi\right)+NST+O\left(\epsilon^{2}\right),\]
 where \emph{NST} stands for the nonsecular terms. As the next step,
we obtain following RG equations:\[
\frac{dA_{R}}{dt}=-\frac{i\epsilon\Omega}{2}B_{R}J_{0}\left(\xi\right)+O\left(\epsilon^{2}\right),\]
 \[
\frac{dB_{R}}{dt}=-\frac{i\epsilon\Omega}{2}A_{R}J_{0}\left(\xi\right)+O\left(\epsilon^{2}\right).\]
 The latter system is readily solved, and with the initial conditions
$c_{g}\left(0\right)=1$, $c_{e}\left(0\right)=0$ produces:\begin{align}
c_{g} & =\cos\left\{ \frac{\epsilon\Omega}{2}J_{0}\left(\xi\right)t\right\} +\epsilon\frac{\Omega}{2}\sin\left\{ \frac{\epsilon\Omega}{2}J_{0}\left(\xi\right)t\right\} \nonumber \\
 & \times\sum_{k=-\infty,k\ne0}^{\infty}J_{k}\left(\xi\right)\frac{\textrm{e}^{-ikt}-1}{k}+O\left(\epsilon^{2}\right),\label{cgtoorderepsilon}\end{align}
\begin{align}
c_{e} & =\sin\left\{ \frac{\epsilon\Omega}{2}J_{0}\left(\xi\right)t\right\} +\epsilon\frac{\Omega}{2}\cos\left\{ \frac{\epsilon\Omega}{2}J_{0}\left(\xi\right)t\right\} \nonumber \\
 & \times\sum_{k=-\infty,k\ne0}^{\infty}J_{k}\left(\xi\right)\frac{1-\textrm{e}^{ikt}}{k}+O\left(\epsilon^{2}\right).\label{cetoorderepsilon}\end{align}
 Note, that coherent destruction of tunneling condition $J_{0}\left(2\xi\right)=0$
\cite{Grossmann1} follows immediately from these results (see also
Ref. \cite{Frasca2}). Continuing the procedure to order $O\left(\epsilon^{2}\right)$,
we obtain \[
\frac{dA_{R}}{dt}=\epsilon^{2}\frac{\Omega^{2}}{4}FA_{R}-\frac{i\epsilon\Omega}{2}B_{R}J_{0}\left(\xi\right),\]
\begin{equation}
\frac{dB_{R}}{dt}=-\frac{i\epsilon\Omega}{2}A_{R}J_{0}\left(\xi\right)-\epsilon^{2}\frac{\Omega^{2}}{4}FB_{R},\label{RGB2}\end{equation}
 with\[
F_{1}=-iJ_{0}\left(\xi\right)\sum_{k=-\infty,k\ne0}^{\infty}\frac{J_{k}\left(\xi\right)}{k}-i\sum_{k=-\infty,k\ne0}^{\infty}\frac{J_{k}^{2}\left(\xi\right)}{k}\]
 Even without solving Eq. (\ref{RGB2}), we observe, that localization
condition holds to order $\epsilon^{3}$. For the recently found third
order corrections see Ref. \cite{Frasca4} and references therein.


\end{document}